\documentstyle[prl,aps,multicol,epsf]{revtex}
\begin{document}                                                        
\draft                                                                  

\title{Inter-site Coulomb interaction and Heisenberg exchange}
\author{R. Eder, J. van den Brink, and G. A. Sawatzky}
\address{Department of Applied and Solid State Physics, 
University of Groningen,\\
9747AG Groningen, The Netherlands}
\date{today}
\maketitle

\begin{abstract}
Based on exact diagonalization results for small clusters we
discuss the effect of inter-site Coulomb repulsion 
in Mott-Hubbard or charge transfers insulators.
Whereas the exchange constant $J$ for direct exchange is 
enhanced by inter-site Coulomb interaction, that for 
superexchange is suppressed. 
The enhancement of $J$ in the single-band models holds
up to the critical value for the charge density wave (CDW)
instability, thus opening the way for large values of $J$. 
Single-band Hubbard models with
sufficiently strong inter-site repulsion to be near
a CDW instability thus may provide physical realizations of
$t$$-$$J$-like models with the `unphysical' parameter ratio 
$J/t$$\approx$$1$.
\end{abstract} 
                                                                               
\begin{multicols}{2}

\begin{figure}
\epsfxsize=\hsize
\vspace{0in}
\hspace{0ex}\epsffile{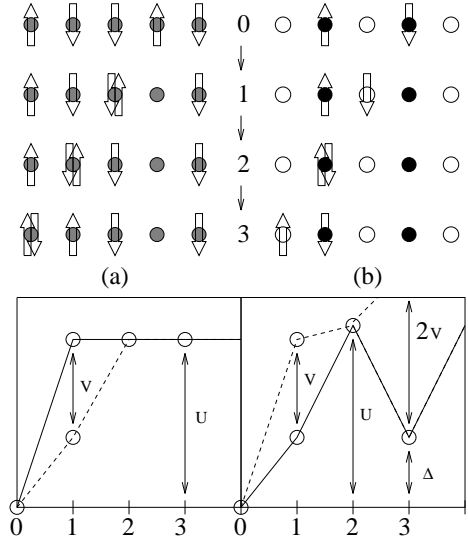}
\vspace{0ex}
\narrowtext
\caption[]{Charge fluctuations in the single band Hubbard model (a)
        and the charge transfer system (b). The dark circles in
        part (b) indicate the `d' sites, the light ones the ligands. The
        lower part of the figure gives the energies of the
        states created by the charge fluctuation with (dashed line)
        and without (full line) a nearest neighbor
        repulsion of strength $V$ taking the hopping integral to be $0$.}
\label{fig1}
\end{figure}

\begin{figure}
\epsfxsize=\hsize
\vspace{0in}
\hspace{0ex}\epsffile{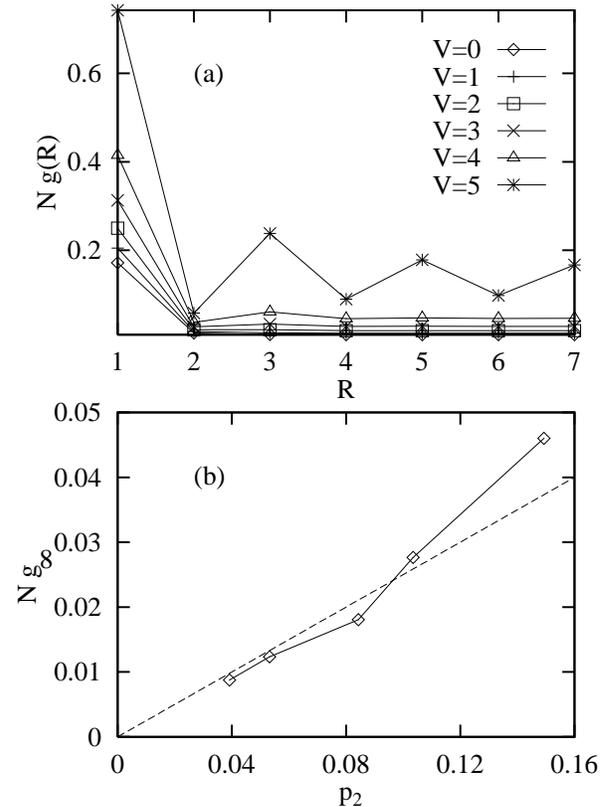}
\vspace{-1.8cm}
\narrowtext
\caption[]{(a) `Empty-double correlation function' $g(R)$ in the ground state
        of a $14$-site 1D Hubbard ring for different values of $V$. 
        The other parameters are $U$$=$$10$, $t$$=$$1$.\\
        (b) Asymptotic value $g_\infty$ vs. $p_2$, the probability
        to have two double occupancies in the ground state.
}
\label{fig2}
\end{figure}

\begin{figure}
\epsfxsize=8cm
\vspace{0cm}
\hspace{0ex}\epsffile{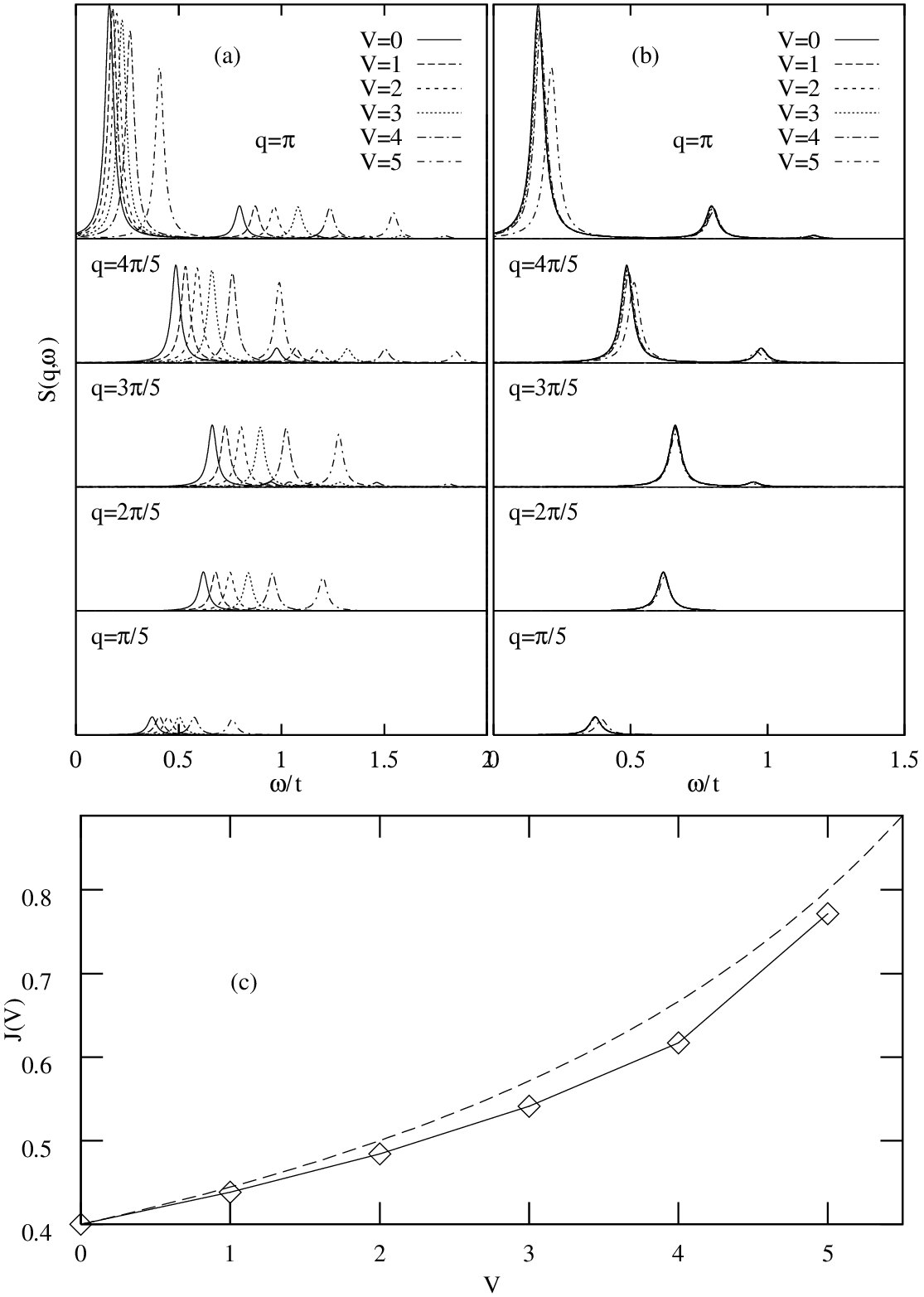}
\vspace{4.0cm}
\narrowtext
\caption[]{(a) Spin correlation functions of a $10$-site 1D Hubbard ring
        for different $V$ ($U$$=$$10$, $t$$=$$1$).\\
        (b) Same as (a) but with the frequencies rescaled so as to
        make the energies of the dominant peak at $2\pi/5$ identical
        for all spectra.\\
        (c) Effective Heisenberg exchange, $J_{eff}(V)$, calculated from
        the rescaling factors used in (b), the dashed line shows
        $J_{eff}(V)= 4t^2/(U-V)$.}
\label{fig3} 
\end{figure}

A frequently used approach to discuss the strong-coupling limit
of Hubbard-type models is the perturbation expansion 
in $t/U$\cite{HarrisLange,ChaoSpalekOles}, which leads to an
effective Hamiltonian operating in the subspace of states without
double occupancies, the well-known $t$$-$$J$ model. By virtue of
its perturbational derivation, the relation $J=4t^2/U$ holds,
so that it seems that a necessary condition
for the validity of $t$$-$$J$-like models is the relation
$J/t= 4(t/U) \ll 1$.
In this work we show that this is not necessarily
the case; rather, Hubbard models with an 
additional inter-site repulsion which is sufficiently strong to
bring the system 
close to a charge-density-wave instability may provide
`physical' realizations of $t$$-$$J$ models with $J/t\approx 1$.
We also show that the situation is reversed in a charge-transfer
insulator, where inter-site Coulomb repulsion on the contrary
suppresses the superexchange. \\
To begin with, it is advantageous to develop a simple picture
of a correlated insulator. 
We restrict ourselves to $1D$ systems in all what follows,
the main reason being that exact diagonalization for reasonably
large $2D$ clusters of the single or even two-band Hubbard model
is not feasible. We expect, however, that the physical considerations
outlined below retain their validity in any dimension.
We consider the single band Hubbard model
\[
H= 
-t \sum_{ i,\sigma} (\;c_{i,\sigma}^\dagger c_{i+1,\sigma}
+ H.c\;)
+ U \sum_i n_{i,\uparrow} n_{i,\downarrow} +
V \sum_{ i} n_i n_{i+1}
\]
where $n_{i,\sigma}$$=$$c_{i,\sigma}^\dagger c_{i,\sigma}$,
$n_i$$=$$n_{i,\uparrow} + n_{i,\downarrow}$.
The model already includes
an additional repulsion between electrons on nearest neighbors $\sim$$V$. 
We also consider the charge tranfer system
\begin{eqnarray}
H &=& 
-t \sum_{ i,j\in N(i),\sigma} ( d_{i,\sigma}^\dagger c_{j,\sigma}
+ H.c)
+ U \sum_i n_{i,\uparrow} n_{i,\downarrow} 
\nonumber \\
&\;&\;\;\;\;\;\;\;-
\Delta \sum_i n_{i} +
V \sum_{i,j\in N(i)} n_{i} n_{j},
\nonumber
\end{eqnarray}
which consists of strongly correlated `$d$'-orbitals separated
by uncorrelated ligands ($N(i)$ denotes the two ligand sites
neighboring the $d$-site $i$).
The on-site energy $-\Delta$ of the $d$-orbitals
is taken to be negative, and in addition to the $d$-site Coulomb repulsion
$U$ we include an additional Coulomb repulsion $V$ between nearest neighbor
$d$ and ligand orbitals.\\
At `half-filling' and in the limit $V$$=0$, $U$$\rightarrow$$\infty$ we 
have precisely one electron per lattice site for the Hubbard model,
and one electron (or hole) per $d$-site in the charge transfer system
(see the states labeled `$0$' in Figure \ref{fig1}).
Switching on the inter-site kinetic energy then
will lead to charge fluctuations: in the Hubbard model, 
an electron can jump to a nearest neighbor, thus creating a `hole'
and a `double occupancy' on nearest neighbors. The energy thereby is
increased by the large amount $U$. In subsequent steps
hole and the double occupancy can separate even further,
but all states generated in this way have the same large Coulomb energy
$U$ (see Figure \ref{fig1}).
For the case of the charge-transfer system, similar
considerations hold, with the energy required to `drag apart'
hole and double occupancy being $min(\Delta, U)$.
By analogy with the problem of a particle moving in a
1D attractive $\delta$-potential one may therefore expect that
the probability to find a hole and a double occupancy at 
a distance of $n$ lattice spacings decreases rapidly with $n$.
The canonical transformation to a Heisenberg or $t$$-$$J$ 
Hamiltonian with only nearest neighbor exchange
then obviously is equivalent to assuming that states where
hole and double occupancy are more distant than 
nearest neighbors (nearest neighbor $d$-sites for the case of the 
charge transfer system) can be completely neglected. \\
Let us now consider the case $V>0$.
For the single-band Hubbard model the energy necessary
to create a charge fluctuation {\em on nearest neighbors} is lowered 
to $U-V$ (see Figure \ref{fig1}). The energy to drag apart
hole and double occupancy, however, remains
unchanged, so that the separation between the two Hubbard
bands (which are associated with `free' propagation of
hole or double occupance)
stays unaffected\cite{comment,Jeroen}. 
Invoking again the analogy with a particle moving in a 1D attractive
potential one might instead expect the formation of a `bound state'
between a hole and a double occupancy on nearest neighbors,
which 
manifests itself as an excitonic excitation 
at an energy well below the Hubbard gap in the optical
conductivity\cite{Jeroen}.
The $V$-term therefore does not reduce the Hubbard gap,
and in particular does not `screen' $U$ to turn the correlated insulator
into a weakly correlated metal. The question remains, however,
whether a transformation to a $t$$-$$J$ model with only 
nearest neighbor exchange still
makes sense in the presence of the $V$-term.
Obviously, if there is an appreciable
probability to have the hole-double occupancy pair
on more distant than nearest neighbor sites, longer range
spin correlations will become important in the effective
low energy Hamiltonian. To address this question we use
exact diagonalization of small clusters; 
we note that for the short-range/high energy
processes which mediate the Heisenberg exchange
finite size effects probably play only a minor role. We
introduce $n_{i,d} = n_{i,\uparrow}  n_{i,\downarrow}$ and
$n_{i,h} = (1-n_{i,\uparrow})(1- n_{i,\downarrow})$, i.e., the
density operators for doubly occupied and empty sites and
study the `empty-double correlation function'
\begin{equation}
g(R) = \frac{1}{N}\sum_i \langle n_{i,h} n_{i+R,d} \rangle
\end{equation}
(where $N$ denotes the chain length)
as a function of the nearest neighbor repulsion $V$.
This is shown in Figure \ref{fig2}.
Obviously $g(R)$ is  peaked
at $R=1$, i.e., independently of the value of $V$ `hole' and `double 
occupancy' strongly
prefer to be on nearest neighbors. For $R>1$ and not too large $V$
the empty-double correlation
immediately levels off to a practically $R$-independent value $g_\infty$;
this can be understood as the contribution from holes and double
occupancies belonging to different `pairs': since the
positions of the pairs are uncorrelated, this component has no more 
$R$-dependence.
To check this interpretation, Fig. \ref{fig2}b
shows $g_{\infty}$ plotted versus $p_2$, the probability to
have precisely two double occupancies in the system
(for our relatively short chain and large value of $U$
the probability to have $n$ double occupancies decays
exponentially with $n$, so that we may safely
neglect corrections due to states
with $3$ or more double occupancies).
To good approximation, $N \cdot g_\infty$
equals the probability to have more than one
double occupancy in the chain, consistent with our interpretation.
Assuming that a Heisenberg antiferromagnet 
with exchange constant $J= 4t^2/(U-V)$ describes the system well,
we can moreover give a simple estimate of $g(R=1)$:
the number of doubly occupied sites is
$\langle N_d \rangle = \partial_{U-V} E_0$, where
$E_0 = -N \cdot (4t^2/(U-V)) 
ln\;2$ is
the ground state energy of the Heisenberg antiferromagnet.
Taking into account that on the average
$50$\% of the double occupancies are to the right of `their' hole
we find $N g(R=1)$$=$$(1/2) \cdot \langle N_d \rangle 
$$=$$ 2 N (t/(U-V))^2\; ln\;2$.
With the parameters
used in Figure \ref{fig2} we find e.g. $Ng(R=1) = 0.1664$ for $V=0$,
$Ng(R=1) = 0.6654$ for $V=5$ which values obviously provide
reasonable estimates.\\
For large values of
$V\approx U/2$, i.e. the critical value for the onset of
the CDW instability in 1D\cite{Hirsch}
we see in addition a modulation of $g(R)$ at longer distances with
a period of two lattice spacings. This
is obviously is a precursor of
the CDW ordering which occurs for larger $V$.
We may thus expect that $V=U/2$ represents a limiting case, where
the mapping to a nearest neighbor-exchange model starts to
become inaccurate.\\
The empty-double correlation function $g(R)$ thus demonstrates 
that the main effect of $V$ is the enhancement of 
{\em nearest neighbor} charge fluctuations.
This suggests that simple nearest neighbor
Heisenberg exchange remains a good approximation also for the
case $V$$>$$0$, the main difference as compared to
the case $V=0$ being the enhancement of charge fluctuations
and hence the Heisenberg exchange $J$. To check this we
consider the dynamical spin correlation function $S(q,\omega)$,
shown in Figure \ref{fig3}a for different values of $V$.
With increasing $V$, all peaks move to
higher energy, the overall shape of the spectra being unaffected.
Rescaling the frequencies
the different $S(q,\omega)$ can be made practically
identical, as demonstrated in Figure \ref{fig3}b.
Assuming the relation $J=4t^2/U$ to hold for $V$$=$$0$,
we can use the rescaling factors to extract the
effective Heisenberg exchange $J_{eff}(V)$,
shown in Figure \ref{fig3}c. Obviously values 
of $J_{eff}/t$, as large as $0.8$, i.e. far outside 
the `perturbational' range are possible. Figure \ref{fig3}c also 
shows the simple estimate $J_{eff}(V) = 4t^2/(U-V)$, which provides
a reasonable approximation for the numerical values.
Let us stress that for small values of $V$ this functional form of
$J_{eff}(V)$ is rather trivial; what is not, however, is the
validity of this relation up to the value $V$$=$$U/2$.\\
We briefly turn to the charge transfer system.
From Figure \ref{fig1} we may expect that the $V$-term
effectively increases the charge transfer energy $\Delta$ to
$\Delta + V$ (it should be stressed, that this refers to
`bare' or `model' parameters; in the case of the charge transfer insulator,
the energy required to drag apart hole and double occupancy
\begin{figure}
\epsfxsize=8cm
\vspace{0cm}
\hspace{0ex}\epsffile{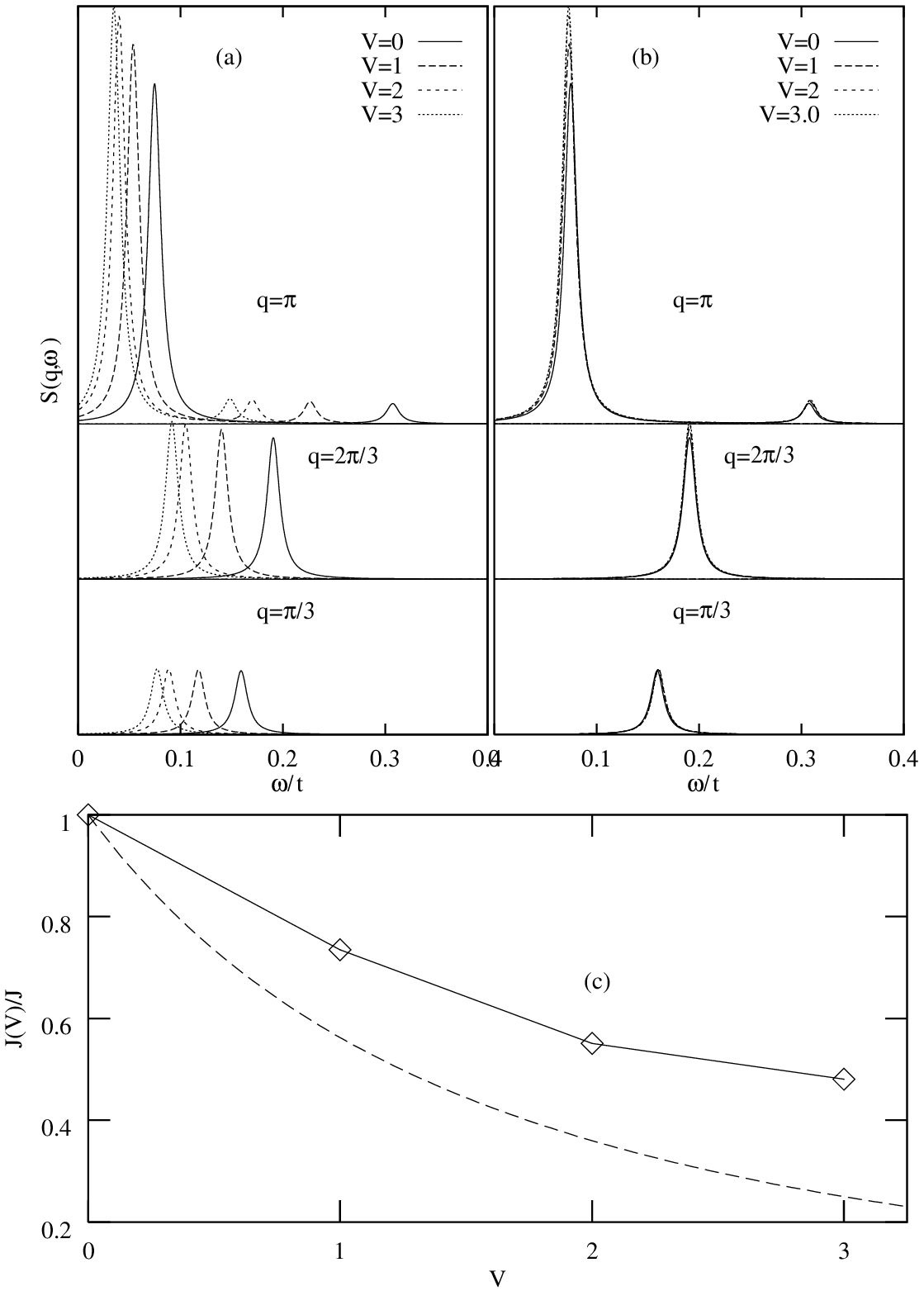}
\vspace{2.2cm}
\narrowtext
\caption[]{(a) Spin correlation functions of a $6$-unit cell 1D Charge transfer
        ring for different $V$ ($\Delta$$=$$3$, $U$$=$$8$, $t$$=$$1$).\\
        (b) Same as (a) but with the frequencies rescaled so as to
        make the energies of the dominant peak at $2\pi/3$ identical
        for all spectra.\\
        (c) Effective Heisenberg exchange, $J_{eff}(V)$, calculated from
        the rescaling factors used in (b). The dashed line has been
        calculated from $4^{th}$ order perturbation theory.
\label{fig4} }
\end{figure}
becomes $min(\Delta+2V, U)$ in the presence of the nearest neighbor
repulsion, 
so that the {\em experimentally measured}
charge transfer gap would be increased to $\Delta+2V$).
Contrary to the case of direct exchange,
this suggests a suppression of the exchange constant $J$.
This is demonstrated in Figure \ref{fig4}, which shows the dynamical spin
correlation function for the charge transfer system
with different values of $V$.
From the rescaling of the SCF (see Figure \ref{fig4}b)  we can 
again extract the effective
exchange constant, which is shown in Figure \ref{fig4}c.
The figure also shows a simple estimate based on $4^{th}$ order
perturbation theory, where 
$J$$=$$(2t_{pd}/(\Delta+V))^2\cdot(1/\Delta +1/U)$,
which however provides only a rough estimate.\\
We return to the single band model, and consider the doped case. 
In this case, the canonical transformation 
gives the strong-coupling model\cite{ChaoSpalekOles}:
\begin{eqnarray}
H &=& -t \sum_{i,\sigma} 
(\; \hat{c}_{i,\sigma}^\dagger \hat{c}_{i+1,\sigma} + H. c.\;) 
+ J \sum_i (\; \vec{S}_i \cdot \vec{S}_j - \frac{n_i n_j}{4}\;)
\nonumber \\
&+& \frac{J}{4} 
\sum_{i,\sigma}\;
[\;( \hat{c}_{i,\sigma}^\dagger n_{i+1,\bar{\sigma}} \hat{c}_{i+2,\sigma} - 
\nonumber \\
&\;&\;\;\;\;\;\;\;\;\;\;\;\;\;\;-
\hat{c}_{i,\sigma}^\dagger 
\hat{c}_{i+1,\bar{\sigma}}^\dagger 
\hat{c}_{i+1,\sigma}
\hat{c}_{i+2,\bar{\sigma}} ) + H.c. \;]
\label{sc}
\end{eqnarray}
One may expect, that the projected model (\ref{sc}) is meaningful
only for the description of the low energy part of the eigenvalue
spectrum - at higher energy the extended Hubbard model may
show additional structure due to the excitonic excitations,
which are not accounted for in the projected model.
\begin{figure}
\epsfxsize=8cm
\vspace{0cm}
\hspace{0ex}\epsffile{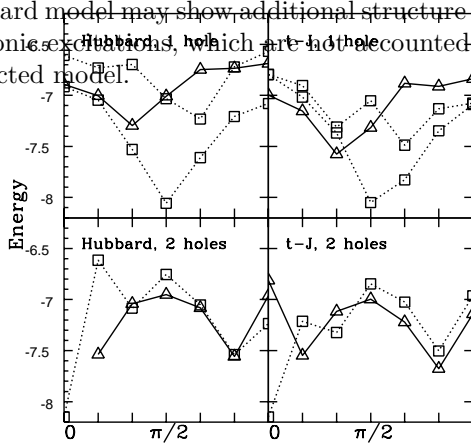}
\vspace{-5cm}
\narrowtext
\caption[]{Lowest eigenvalues 
        for all allowed momenta of $12$-site rings of extended
        Hubbard model ($U$$=$$10$, $V$$=$$5$)
        and extended $t-J$ model ($J/t=0.8$) for $11$ and $10$ electrons
        (i.e. one and two holes in the half-filled band).
        For the single hole, squares (triangles) denote states with
        $S=1/2$ ($S=3/2$), for two holes states with $S=0$ ($S=1$).
        States of like spin have been grouped into `bands' only 
        to guide the eye.}
\label{fig5}
\end{figure}
Figure \ref{fig5} therefore compares the energies of the lowest
eigenvalues of the extended Hubbard model with
$U/V$$=$$2$, and the strong coupling model
with $J/t=0.8$ (we have used the spin correlation function at
half-filling to `gauge' the value of $J$; moreover, a nearest
neighbor density repulsion with $V=5$ has been added to the strong
coupling model).
While the agreement between the two level schemes
is not really pefect, there is an obvious correspondence.
Deviations are largest for the $2$ hole states with
$S$$=$$0$ near $\pi/3= 4k_F$, which probably indicates a somewhat higher
`holon' velocity in the actual Hubbard model.
This may be related to the fact that
$U/V=2$ is at the very borderline of stability against formation
of a CDW\cite{Hirsch}, and the empty-double correlation function showed 
already some signature of the CDW.
Apart from that, the strong coupling model 
reproduces the overall energy scales of the
Hubbard model quite well, which indicates
that for low energy excitations,
despite its apparently unphysically
large exchange constant, the strong coupling model
may indeed provide the appropriate 
`effective' model.\\ 
In summary, we have investigated the influence of inter-site
Coulomb repulsion in the Hubbard model.
For single band Hubbard models, our results suggest an enhancement
of the exchange constant, roughly as $J(V)= 4t^2/(U-V)$, whereas
the spin dynamics still remains consistent with
that of an ideal Heisenberg antiferromagnet. The remarkable
feature of the results is that this rescaling seems to hold
up to the CDW instability, which opens the way to
unexpectedly large values of the ratio $J/t$.
In particular, for the strong correlation case close to
a CDW instability in 1D, our results suggest values of $J$ 
comparable to the original hopping integral. 
Therefore, real materials which may be described
by a strong correlation single band Hubbard model close to a CDW instability
may provide realizations of $t$$-$$J$-like models with `unphysically' large
values of $J/t\approx 1$.
For higher spatial dimensions, the critical value 
of $U/V$ for the formation of the CDW state is smaller than in 1D,
which seems to preclude large $J$ values due to inter-site repulsion.
However, an alternative possibility here could be the suppression of the
CDW due to a non-bipartite lattice structure, which can
rule out the CDW even for large values of $V$; an interesting example
here could be solid $C_{60}$, which crystallizes in the $fcc$ structure.
Auger spectroscopy\cite{Lof} gives values of $U\approx 1.5eV$, whereas
analysis of the exciton dispersion\cite{AnnaMaria}
in undoped $C_{60}$ suggests values of $U-V$ as small as
$0.35eV$, which is also consistent with other experimental 
estimates\cite{Bruehwiler}.
This is to be compared to the average bandwidth
of the LUMO-derived bands of $\approx 0.3 eV$\cite{Satpathy}.
While the inter-site
Coulomb repulsion may be more efficiently screened
in the doped material than the on-site $U$, leading to an increase of $U-V$,
one still may expect relatively large values of $J$.\\

\end{multicols}

\end{document}